\documentclass{ws-ijmpd}
\usepackage[super,compress]{cite}
\begin{document}


\markboth{Deng Wang}
{Cosmology with Type II Supernovae}

%
\catchline{}{}{}{}{}
%

\title{Cosmology with Type II Supernovae}

\author{Deng Wang}

\address{{Department of Astronomy, School of Physics of Astronomy, Shanghai Jiao Tong University, Shanghai 200240, China},
cstar@mail.nankai.edu.cn}

\maketitle

\begin{history}
\received{Day Month Year}
\revised{Day Month Year}
\end{history}

\begin{abstract}
With the recent progresses on the Type II supernovae, we attempt to investigate whether there does exist new physics beyond the standard cosmological paradigm, i.e., the cosmological constant $\Lambda$ plus cold dark matter ($\Lambda$CDM). Constraining four alternative cosmological models with a data combination of currently available Type II supernovae calibrated by the standard color method, Type Ia supernovae, baryon acoustic oscillations, cosmic microwave background and cosmic chronometers, at the $1\sigma$ confidence level, we find that: (i) a spatially flat universe is supported for the non-flat $\Lambda$CDM model; (ii) the constrained equation of state of dark energy $\omega$ is consistent with the $\Lambda$CDM hypothesis for the $\omega$CDM model where $\omega$ is a free parameter; (iii) for the decaying vacuum model, there is no evidence of interaction between dark matter and dark energy in the dark sector of the universe; (iv) there is also no hint of dynamical dark energy for the dark energy density-parametrization scenario. It is very obvious that a larger Type II supernovae sample is required, if we expect to draw definitive conclusions about the formation and evolution of the universe.
\end{abstract}


\ccode{PACS numbers:}


\section{Introduction}
It has been for two decades since Type Ia supernovae (SNe Ia), which plays a crucial role in revealing the evolution and composition of the universe, led to a revolutionary discovery that our universe is during a phase of accelerated expansion \cite{1,2,3}. This mysterious accelerated phenomenon has been named as dark energy (DE), a new matter source, by modern cosmologists. During the past two decades, the existence of DE has also been confirmed by many independent cosmological probes such as the cosmic microwave background (CMB) radiation \cite{4,5}, baryon acoustic oscillations (BAO) \cite{6,7}, weak gravitational lensing \cite{8}, X-ray clusters \cite{9,10} and superluminous SNe \cite{11}. Since the SNe Ia cosmology reaches a mature state to date and the nature of DE is still unclear for us, it is essential to develop other independent distance indicators to constrain the DE.

One of the most promising independent probes to explore the evolution of the late universe is the Type II supernovae (SNe II), which are generally 1-2 mag fainter than SNe Ia and characterized by a plateau of varying steepness in their light curves \cite{12} and the presence of strong hydrogen features in their spectra (see \cite{13,14} for details). Up to now, there are five main standardized methods to calibrate the SNe II: (i) expanding photosphere method (EPM) \cite{15}; (ii) standard candle method (SCM) \cite{16}; (iii) spectral-fitting expanding atmosphere method (SEAM) \cite{17,18}; (iv) a generalized version of SCM, photospheric magnitude method (PMM) \cite{19}; (v) photometric color method (PCM) \cite{20}. Recently, to re-assess the utility of SNe II as distance indicators, Gall {\it et al.} \cite{21} present photometry and spectroscopy of nine new SNe II-P/L lying in the redshift range $0.045\lesssim z \lesssim 0.335$ and exhibit an updated SNe II Hubble diagram. After applying the EPM and SCM to each target, they find that both methods yields distances which are in reasonable agreement with each other. Subsequently, they also find that using the Hubble-flow SNe II-L as distance indicators can produce similar distances as the SNe II-P. By combining data from three different surveys: the Carnegie Supernovae Project-I (CSP-I), Sloan Digital Sky Survey II (SDSS-II) and Supernovae Legacy Survey (SNLS),  Jaeger {\it et al.} \cite{22} construct another SNe II Hubble diagram. After applying the PCM and SCM to this new data sample, they just obtain weak constraints on the cosmological parameters. Interestingly, in the framework of standard cosmological model, i.e., $\Lambda$CDM, they find the present-day matter density ratio $\Omega_m=0.32^{+0.30}_{-0.21}$ which provides a new independent evidence for DE at nearly 2$\sigma$ confidence level (CL). Furthermore, in a follow-up study \cite{23}, they construct the highest-redshift SNe II Hubble diagram to date by applying the SCM to SN2016jhj ($z=0.3398\pm0.0002$) from the Hyper Suprime-Cam (HSC) survey.

It is noteworthy that the above previous works \cite{21,22,23} just consider the case of $\Lambda$CDM cosmology. Although the $\Lambda$CDM model has been verified to be very successful in describing various cosmological phenomena from the origin and evolution of large scale structure to the late-time acceleration, it is not perfect and faces at least two great challenges \cite{24}: (i) Why is the value of $\Lambda$ unexpectedly small with respect to any physically meaningful scale, except the current horizon scale ? (ii) Why this value is not only small, but also surprisingly close to another unrelated physical quantity, $\Omega_m$ ? This implies that the $\Lambda$CDM model may not be the true one governing the formation and evolution of the universe. Along this logical line, we are motivated by exploring whether there exists new physics beyond the standard cosmological model in light of new SNe II data. Specifically, we combine 61 SNe II data points from \cite{23} with SNe Ia, BAO, CMB and cosmic chronometers (CC) to constrain four alternative cosmological models, and do not find
the hints of new physics.

This study is structured in the following manner. In the next section, we give a brief introduction to four cosmological models to be constrained by the SNe II data. In Section III, we introduce our analysis methodology and describe the SNe II data sample we use. In Section IV, we exhibit
our numerical analysis results. In the final section, the discussions and conclusions are presented.

\section{Models}
In this section, we briefly review four alternative cosmological models to be constrained by data. In a Friedmann-Robertson-Walker (FRW) universe under
the framework of general relativity (GR), we investigate these models and just focus on the late universe, consequently neglecting the contribution
from radiation in the cosmic pie. Starting from the well-known Friedmann equations, we derive the dimensionless Hubble parameter (DHP) for the $\Lambda$CDM model as
\begin{equation}
E_{\mathrm{\Lambda CDM}}(z)=\left[\Omega_{m}(1+z)^3+1-\Omega_{m}\right]^{\frac{1}{2}}, \label{1}
\end{equation}
while for the non-flat $\Lambda$CDM (o$\Lambda$CDM) model it can be written as
\begin{equation}
E_{o\mathrm{\Lambda CDM}}(z)=\left[\Omega_{m}(1+z)^3+\Omega_{k}(1+z)^2+1-\Omega_{m}-\Omega_{k}\right]^{\frac{1}{2}}, \label{2}
\end{equation}
where $\Omega_{k}$ denotes the present-day curvature density ratio parameter.

To study the basic state of DE, we place the constraint on the equation of state (EoS) of DE. Specifically, we consider the simplest parameterized form $\omega(z)=\omega$, namely the $\omega$CDM model, where the DE is a single negative cosmic fluid homogeneously permeating in the universe. The corresponding DHP for a spatially flat ¦ØCDM model is shown as
\begin{equation}
E_{\mathrm{\omega CDM}}(z)=\left[\Omega_{m}(1+z)^3+(1-\Omega_{m})(1+z)^{3(1+\omega)}\right]^{\frac{1}{2}}. \label{3}
\end{equation}

For a long time, an important problem in modern cosmology is whether or not there is interaction between DM and DE in the dark sector of the universe. To investigate this topic, we constrain a popular decaying vacuum (DV) model proposed by Wang and Meng \cite{25}, and its DHP is conveniently expressed as
\begin{equation}
E_{\mathrm{DV}}(z)=\left[\frac{3\Omega_{m}}{3-\epsilon}(1+z)^{3-\epsilon}+1-\frac{3\Omega_{m}}{3-\epsilon}\right]^{\frac{1}{2}},   \label{4}
\end{equation}
where $\epsilon$ represents the typical free parameter of this DV model. It is noteworthy that $\epsilon$ means a small modified matter expansion rate. $\epsilon<0$ implies that the momentum transfers from DM to DE and vice versa.

Another puzzling and important problem is whether the DE evolves over time or not. To address this issue easily, we take a DE density-parametrization (DEDP) model proposed by us \cite{26} into account, and its corresponding DHP shall be written as
\begin{equation}
E_{\mathrm{DEDP}}(z)=\left[\Omega_{m}(1+z)^{3}+(1-\Omega_{m})(1+\delta-\frac{\delta}{1+z})\right]^{\frac{1}{2}},   \label{5}
\end{equation}
where $\delta$ is a free parameter characterizing the evolution behavior of this DEDP model. It is not difficult to see that this model reduces to the $\Lambda$CDM case when $\delta=0$, and that if $\delta$ has any departure from zero, the DE will be dynamical.

\section{Methodology and Data}
The SCM, which is the most used to standardize SNe II, is based on the correlation between the photospheric expansion velocity and the intrinsic luminosity. In this study, to constrain the nature of DE, we apply the SCM to standardize SNe II. More specifically, we utilize two corrections standardize the observed magnitude of SNe II: the expansion velocity, and the color correction that accounts for host-galaxy extinction. Therefore, the observed magnitude of SNe II is modeled as
\begin{equation}
m_i^{model}=\mathcal{M}_i-\alpha \mathrm{log}_{10}\left( \frac{\nu_{\mathrm{H\beta}}}{<\nu_{\mathrm{H\beta}}> \mathrm{km\, s}^{-1}} \right)+\beta(r-i)+5\mathrm{log}_{10}[\mathcal{D}_L(z_{\mathrm{CMB}}|\vec{\theta})],   \label{6}
\end{equation}
where $(r-i)$ denotes the color correction, $\mathcal{D}_L(z_{\mathrm{CMB}}|\vec{\theta})=H_0d_L$ is the cosmology-dependent luminosity distance, $H_0$ is the Hubble constant, $z_{\mathrm{CMB}}$ is the CMB redshift, $\vec{\theta}$ is the cosmological parameter vector (e.g., for the DEDP case, $\vec{\theta}=(\Omega_m, \delta)$), and $\alpha$, $\beta$ and $\mathcal{M}_i$ are free parameters to be constrained by data.

Due to the lack of the SNe II with an accurate distance estimation in the currently available sample, we just consider the relative distances and define the $H_0$-free absolute magnitude as $\mathcal{M}_i=M_i-5\mathrm{log}_{10}(H_0)+25$ as done in the origin work \cite{3}. It is worth noting that we center the expansion velocity and color correction by using the mean H$\beta$ $\lambda$4861 velocity $<\nu_{\mathrm{H\beta}}>\approx5910\, \mathrm{km\, s}^{-1}$ and the mean color correction $<(r-i)>\approx-0.02 \, \mathrm{mag}$ of the entire sample, respectively.

To carry out the standard Bayesian analysis, we shall show the likelihood function of the SNe II data as
\begin{equation}
-2\mathrm{ln}\mathcal{L}=\sum_n\left[ \frac{(m_i^{model}-m_i^{obs})^2}{\sigma^2_{tot}}+2\mathrm{ln}\sigma_{tot} \right] ,\label{7}
\end{equation}
where $m_i^{model}$ denotes the predicted magnitude of a specific cosmological model (see Eqs. (\ref{1}-\ref{5})), $m_i^{obs}$ is the observed magnitude corrected for AKS, the total error of the corresponding model can be expressed as
\begin{equation}
\sigma_{tot}=\left\{\sigma_{obs}^2+\sigma_{m_i}^2+[\beta{\sigma_{(r-i)}}]^2+\left(  \frac{\alpha}{\mathrm{ln}10}\frac{\sigma_{\nu_{\mathrm{H}\beta}}}{\nu_{\mathrm{H}\beta}} \right)+\left[  \frac{5}{\mathrm{ln}10} \frac{\sigma_z(1+z)}{z(1+\frac{z}{2})} \right]^2 \right\}^{\frac{1}{2}}.\label{8}
\end{equation}
It is worth noting that the term $\sigma_{obs}^2$ corresponds to the realistic intrinsic scatter in light of SCM in the Hubble diagram and any misestimates of the expansion velocity, photometric, or redshift errors.

To implement tight constraints on cosmological parameters, we also include SNe Ia, BAO, CMB and CC in our data analysis. These datasets can be described in the following compact manner.

SNe Ia: We use relatively large SNe Ia compilation `` Joint Light-curve Analysis '' (JLA) \cite{JLA} including SNLS, SDSS, HST and several low-$z$ SNe Ia. This sample consists of 740 data points covering the redshift range $z\in[0.01, 1.3]$.

BAO: We adopt BOSS DR12 dataset at three effective redshifts $z_{eff} = 0.38$, 0.51, 0.61 \cite{BAO}.

CMB: We employ the compressed CMB data obtained in \cite{CMB} through a combination of Planck temperature data, Planck lensing and WMAP polarization.

CC: The CC data determined by using the most massive and passively evolving galaxies based on the `` galaxy differential age '' method is model-independent. We utilize the latest 31 data points \cite{CC1,CC2} to constrain cosmological models.

\begin{figure}
\centering
\includegraphics[scale=0.4]{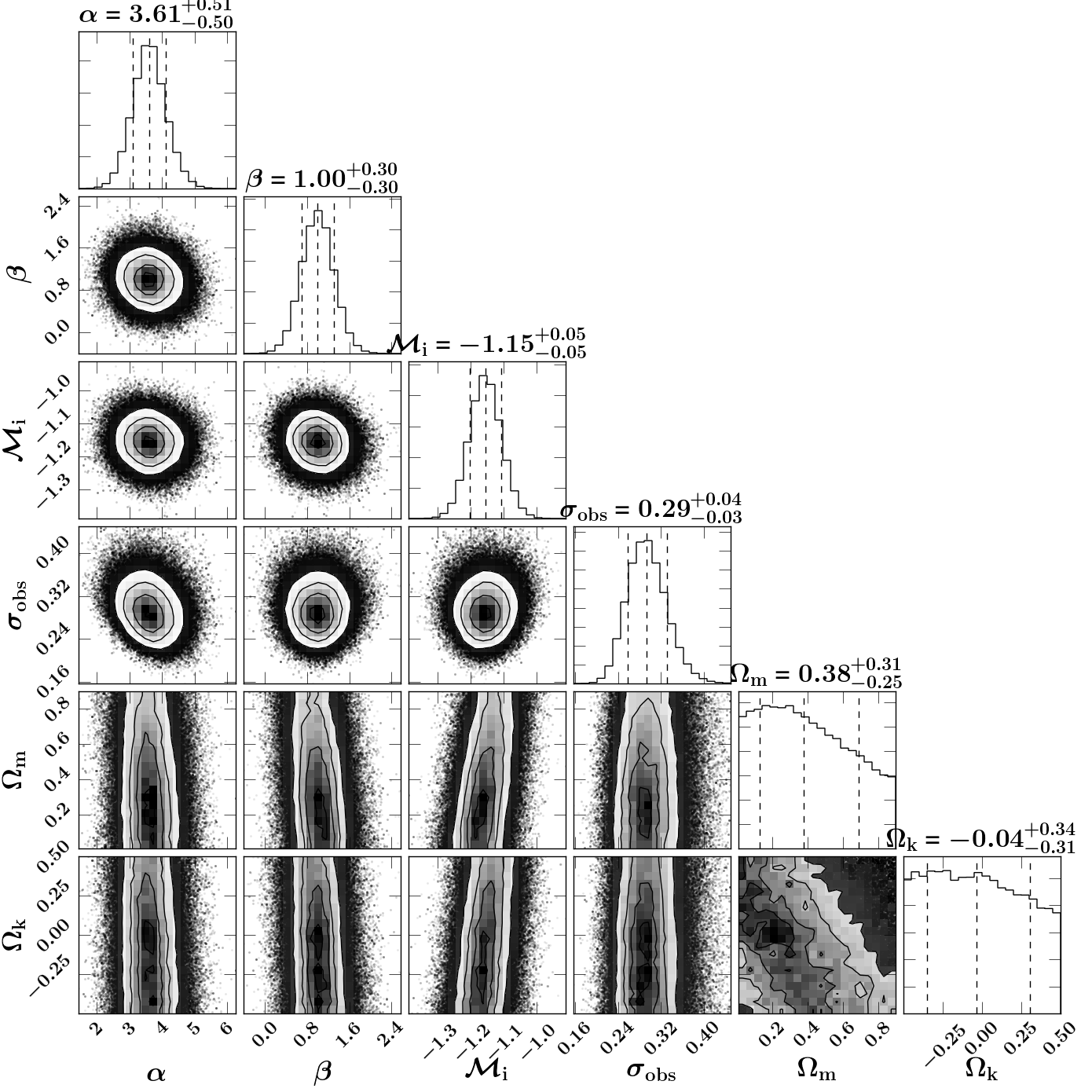}
\caption{Using the currently available SNe II data, we present the 1-dimensional posterior distributions on the individual parameters and $2$-dimensional marginalized contours of the o$\Lambda$CDM model. The $68\%$ confidence ranges of different cosmological parameters are also exhibited in the corner plot.}\label{f1}
\end{figure}
\begin{figure}
\centering
\includegraphics[scale=0.4]{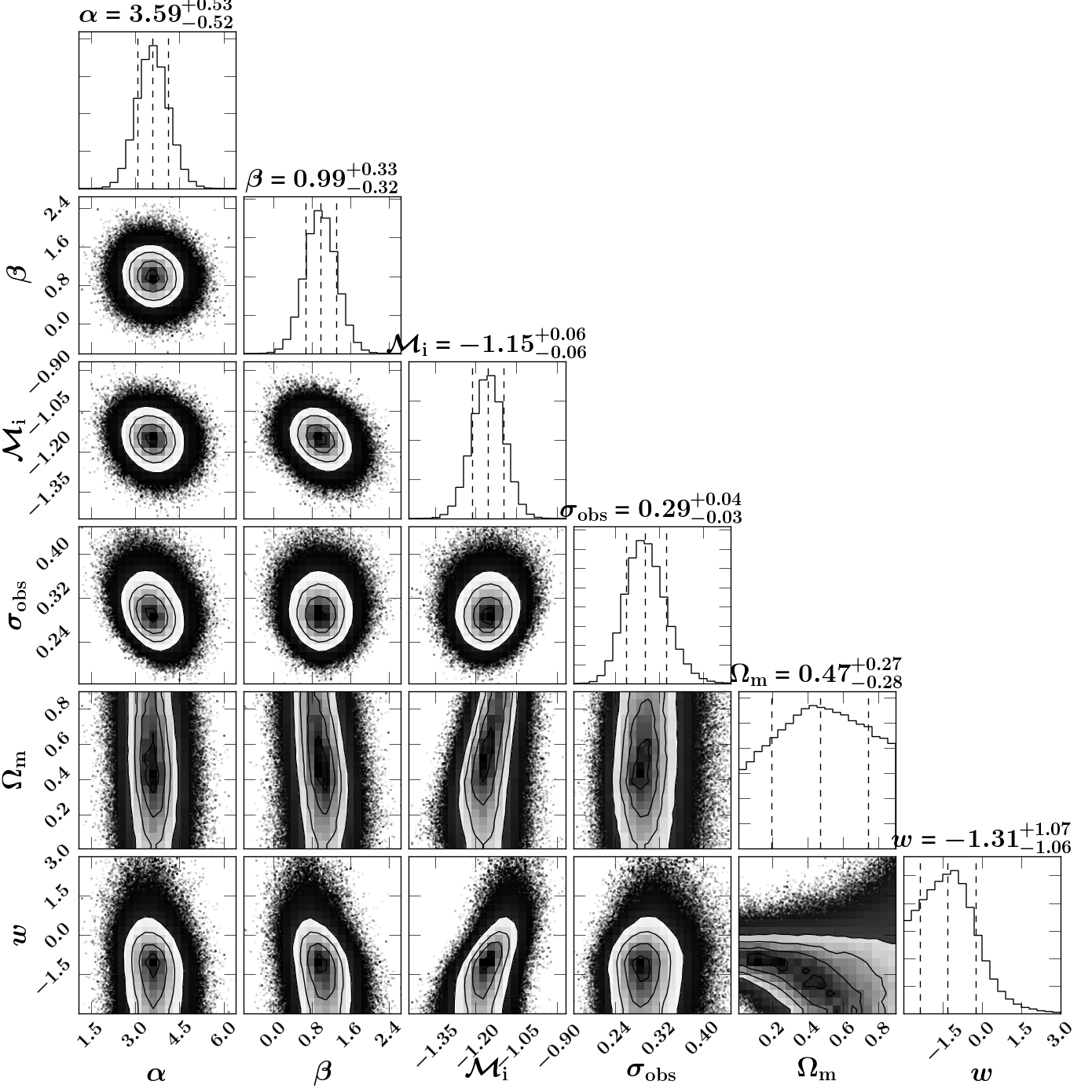}
\caption{Using the currently available SNe II data, we present the 1-dimensional posterior distributions on the individual parameters and $2$-dimensional marginalized contours of the $\omega$CDM model. The $68\%$ confidence ranges of different cosmological parameters are also exhibited in the corner plot. }\label{f2}
\end{figure}
\begin{figure}
\centering
\includegraphics[scale=0.4]{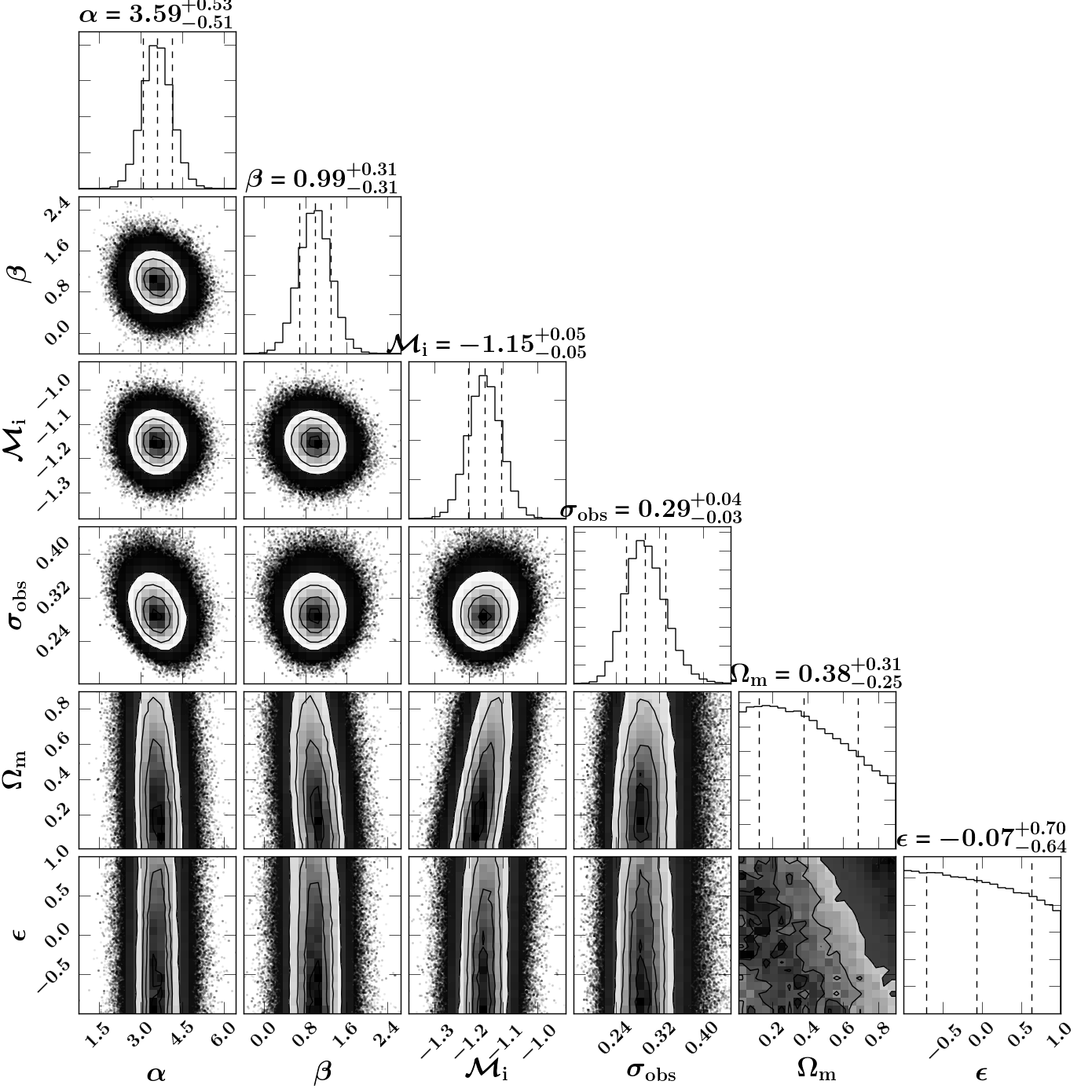}
\caption{Using the currently available SNe II data, we present the 1-dimensional posterior distributions on the individual parameters and $2$-dimensional marginalized contours of the DV model. The $68\%$ confidence ranges of different cosmological parameters are also exhibited in the corner plot.}\label{f3}
\end{figure}
\begin{figure}
\centering
\includegraphics[scale=0.4]{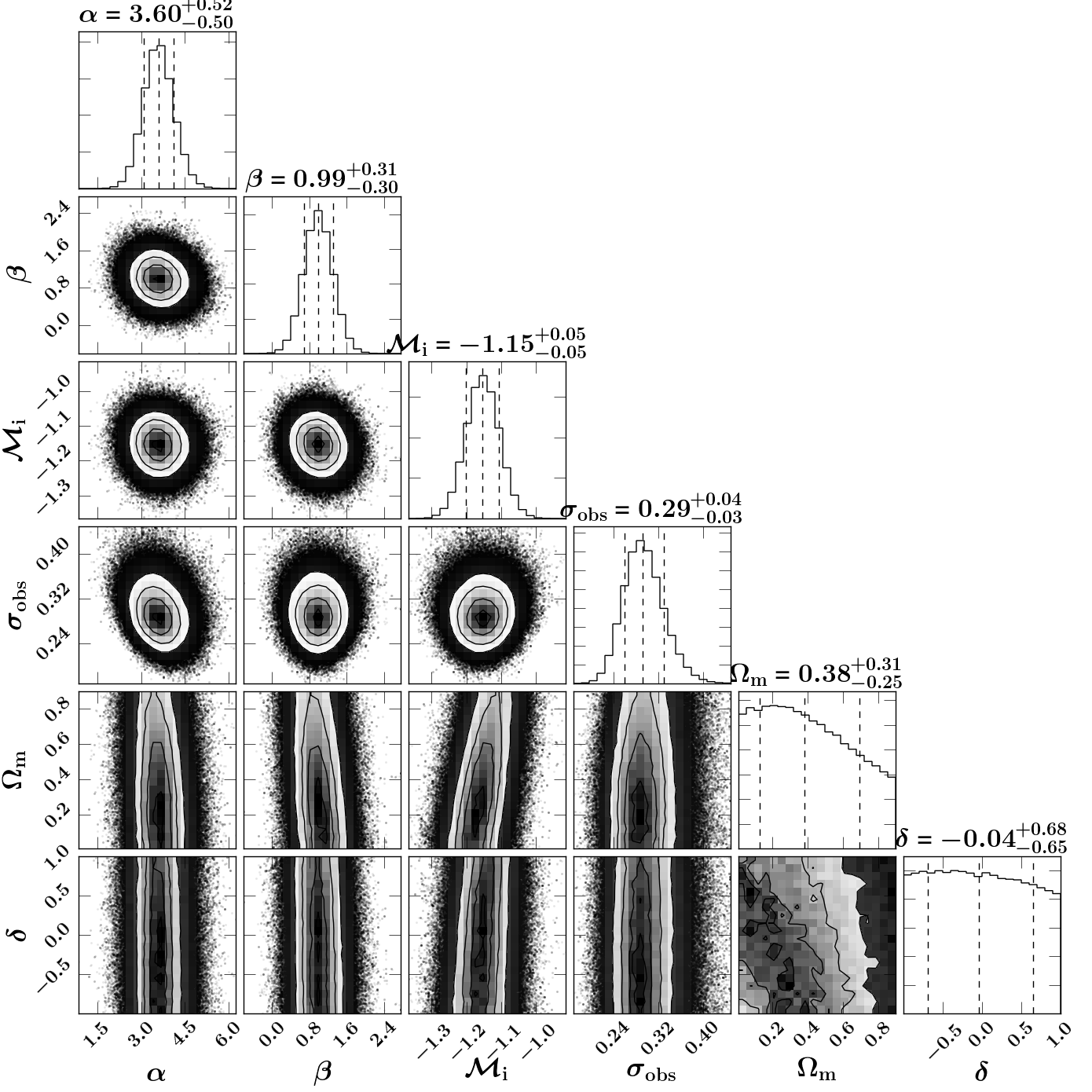}
\caption{Using the currently available SNe II data, we present the 1-dimensional posterior distributions on the individual parameters and $2$-dimensional marginalized contours of the DEDP model. The $68\%$ confidence ranges of different cosmological parameters are also exhibited in the corner plot.}\label{f4}
\end{figure}

\begin{table}[ph]
\renewcommand\arraystretch{1.5}
\tbl{1$\sigma$ confidence ranges of free parameters of five different cosmological models by using SNe II alone.}
{\begin{tabular}{@{}cccccc@{}} \toprule
Parameters            &$\Lambda$CDM          & o$\Lambda$CDM   &$\omega$CDM         &DV           &DEDP                     \\ \colrule
$\Omega_{m}$        &$0.41^{+0.31}_{-0.27}$ &$0.38^{+0.31}_{-0.25}$  &$0.47^{+0.27}_{-0.28}$ &$0.38^{+0.31}_{-0.25}$ &$0.38^{+0.31}_{-0.25}$                           \\
$\Omega_k$                  &---                &$-0.04^{+0.34}_{-0.31}$                      &---            &---             &---                         \\
$\omega$             &---                        &---                      &$-1.31^{+1.07}_{-1.06}$           &---             &---                          \\
$\epsilon$           &---                        &---                      &---            &$-0.07^{+0.70}_{-0.64}$             &---                          \\
$\delta$                  &---                        &---                      &---            &---             &$-0.04^{+0.68}_{-0.65}$                         \\
\botrule
\end{tabular}
\label{t1}}
\end{table}

\begin{table}[ph]
\renewcommand\arraystretch{1.5}
\tbl{1$\sigma$ confidence ranges of free parameters of five different cosmological models by using SNe Ia alone.}
{\begin{tabular}{@{}cccccc@{}} \toprule
Parameters            &$\Lambda$CDM            & o$\Lambda$CDM      &$\omega$CDM           &DV           &DEDP                     \\ \colrule
$\Omega_{m}$        &$0.296^{+0.035}_{-0.034}$ &$0.284^{+0.045}_{-0.050}$  &$0.187^{+0.093}_{-0.087}$ &$0.311^{+0.057}_{-0.052}$ &$0.293^{+0.060}_{-0.070}$                           \\
$\Omega_k$                  &---                        &$0.027^{+0.117}_{-0.082}$                      &---            &---             &---                       \\
$\omega$             &---                        &---                      &$-0.766^{+0.106}_{-0.153}$           &---             &---                          \\
$\epsilon$           &---                        &---                      &---            &$0.1885^{+0.5318}_{-0.4394}$             &---                          \\
$\delta$                  &---                        &---                      &---            &---             &$0.0583^{+0.5317}_{-0.4412}$                         \\
\botrule
\end{tabular}
\label{t2}}
\end{table}

\begin{table}[ph]
\renewcommand\arraystretch{1.5}
\tbl{1$\sigma$ confidence ranges of free parameters of five different cosmological models by using SNe II $+$ SNe Ia.}
{\begin{tabular}{@{}cccccc@{}} \toprule
Parameters            &$\Lambda$CDM           & o$\Lambda$CDM      &$\omega$CDM            &DV           &DEDP                     \\ \colrule
$\Omega_{m}$        &$0.288^{+0.033}_{-0.030}$ &$0.284^{+0.043}_{-0.039}$  &$0.274^{+0.066}_{-0.074}$ &$0.294^{+0.041}_{-0.035}$ &$0.298^{+0.048}_{-0.057}$                           \\
$\Omega_k$                  &---                        &$0.027^{+0.089}_{-0.072}$                      &---            &---             &---                         \\
$\omega$             &---                        &---                      &$-0.976^{+0.145}_{-0.101}$           &---             &---                          \\
$\epsilon$           &---                        &---                      &---            &$0.0399^{+0.2090}_{-0.1354}$             &---                          \\
$\delta$                  &---                        &---                      &---            &---             &$0.0235^{+0.2409}_{-0.2205}$                         \\
\botrule
\end{tabular}
\label{t3}}
\end{table}

\begin{table}[ph]
\renewcommand\arraystretch{1.5}
\tbl{1$\sigma$ confidence ranges of free parameters of five different cosmological models by using SNe II $+$ SNe Ia $+$ BAO $+$ CMB $+$ CC.}
{\begin{tabular}{@{}cccccc@{}} \toprule
Parameters            &$\Lambda$CDM           & o$\Lambda$CDM      &$\omega$CDM           &DV           &DEDP                     \\ \colrule
$\Omega_{m}$        &$0.277^{+0.011}_{-0.009}$ &$0.280^{+0.009}_{-0.009}$  &$0.278^{+0.008}_{-0.008}$ &$0.276^{+0.010}_{-0.011}$ &$0.277^{+0.010}_{-0.011}$                           \\
$\Omega_k$                  &---                        &$-0.009^{+0.011}_{-0.011}$                     &---            &---             &---                        \\
$\omega$             &---                        &---                      &$-1.019^{+0.028}_{-0.027}$           &---             &---                          \\
$\epsilon$           &---                        &---                      &---            &$0.0032^{+0.0093}_{-0.0082}$             &---                          \\
$\delta$                  &---                        &---                      &---            &---             &$0.0081^{+0.0333}_{-0.0539}$                         \\
\botrule
\end{tabular}
\label{t4}}
\end{table}

In order to perform conveniently Bayesian parameter estimation, we employ the online package \textbf{EMCEE} \cite{27}, which is an extensible, a pure-python implementation of Goodman and Weare's Affine Invariant Markov chain Monte Carlo (MCMC) Ensemble sampler. Furthermore, we choose the prior ranges for different cosmological parameters in the following manner: $\alpha$, $\beta$, $\mathcal{M}_i\neq0$, $\sigma_{obs}\in [0.1, 0.45]$, $\Omega_m\in[0.01, 0.9]$, $\Omega_k\in[-0.5, 0.5]$, $\omega\in[-3,3]$, $\epsilon\in[-1,1]$ and $\delta\in[-1,1]$.

In this study, we use 61 SNe II data points compiled in \cite{23} (see Tab. A1), which are selected from four different surveys: 39 from CSP-I, 16 from SDSS-II, 5 from SNLS and 1 from HSC. For more details about the SNe II data reduction procedures and different surveys, the reader is referred to \cite{22,28}. We shall divide our analysis into 4 classes: SNe II, SNe Ia, SNe II $+$ SNe Ia and SNe II $+$ SNe Ia $+$ BAO $+$ CMB $+$ CC.

\section{Analysis results}
\begin{figure}
\centering
\includegraphics[scale=0.4]{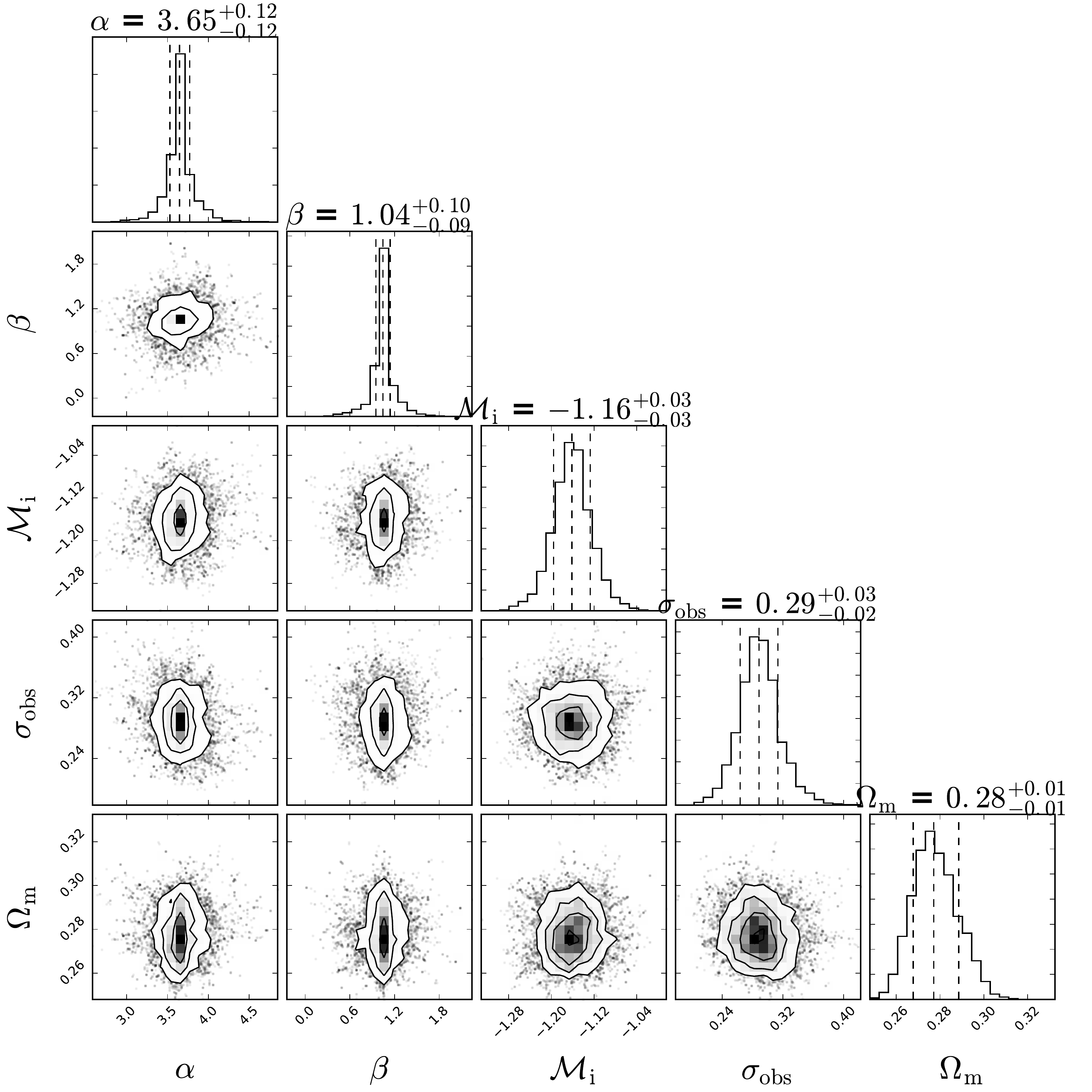}
\caption{Using a data combination of SNe II $+$ SNe Ia $+$ BAO $+$ CMB $+$ CC, we present the 1-dimensional posterior distributions on the individual parameters and $2$-dimensional marginalized contours of the $\Lambda$CDM model. The $68\%$ confidence ranges of different cosmological parameters are also exhibited in the corner plot.}\label{f5}
\end{figure}
\begin{figure}
\centering
\includegraphics[scale=0.4]{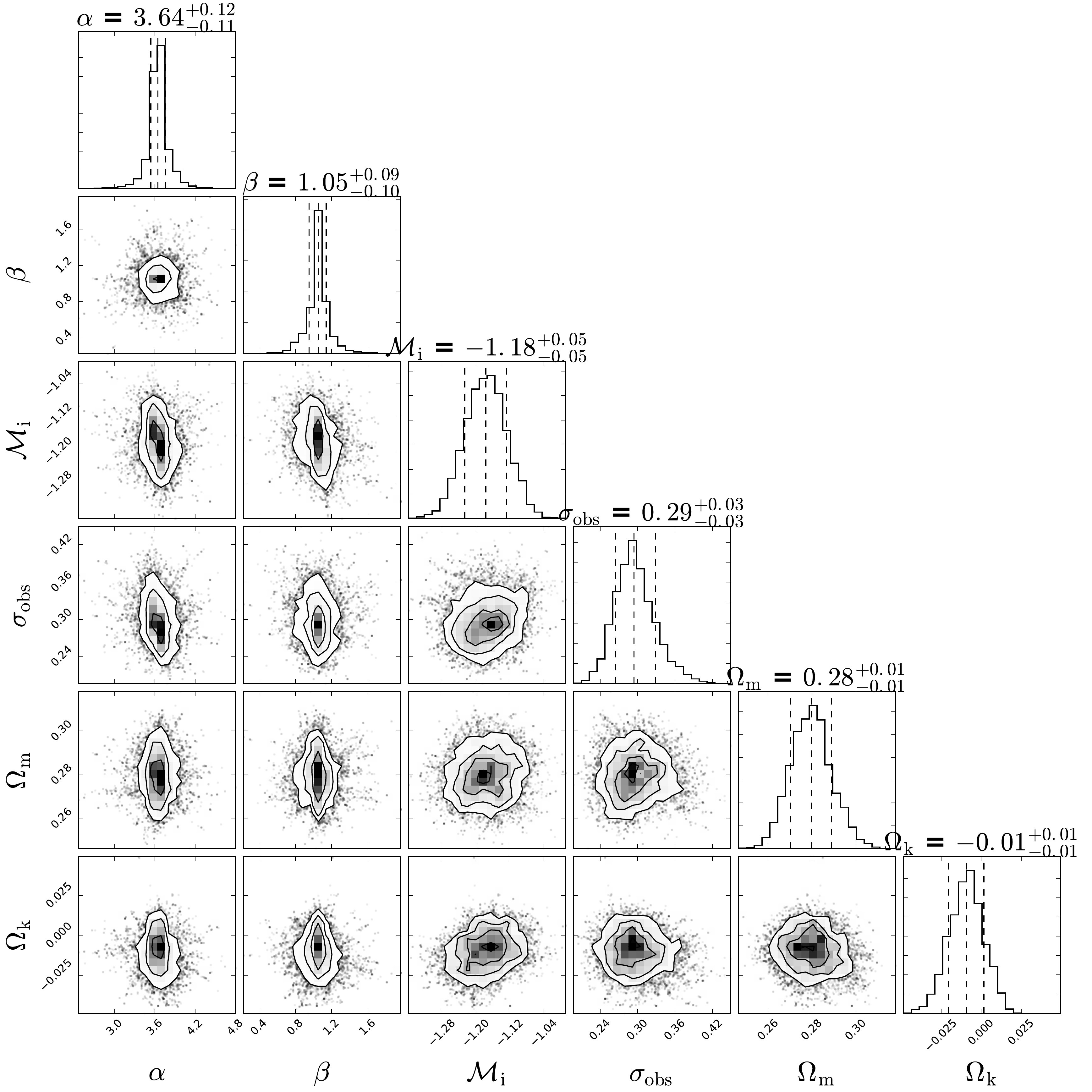}
\caption{Using a data combination of SNe II $+$ SNe Ia $+$ BAO $+$ CMB $+$ CC, we present the 1-dimensional posterior distributions on the individual parameters and $2$-dimensional marginalized contours of the o$\Lambda$CDM model. The $68\%$ confidence ranges of different cosmological parameters are also exhibited in the corner plot.}\label{f6}
\end{figure}
\begin{figure}
\centering
\includegraphics[scale=0.4]{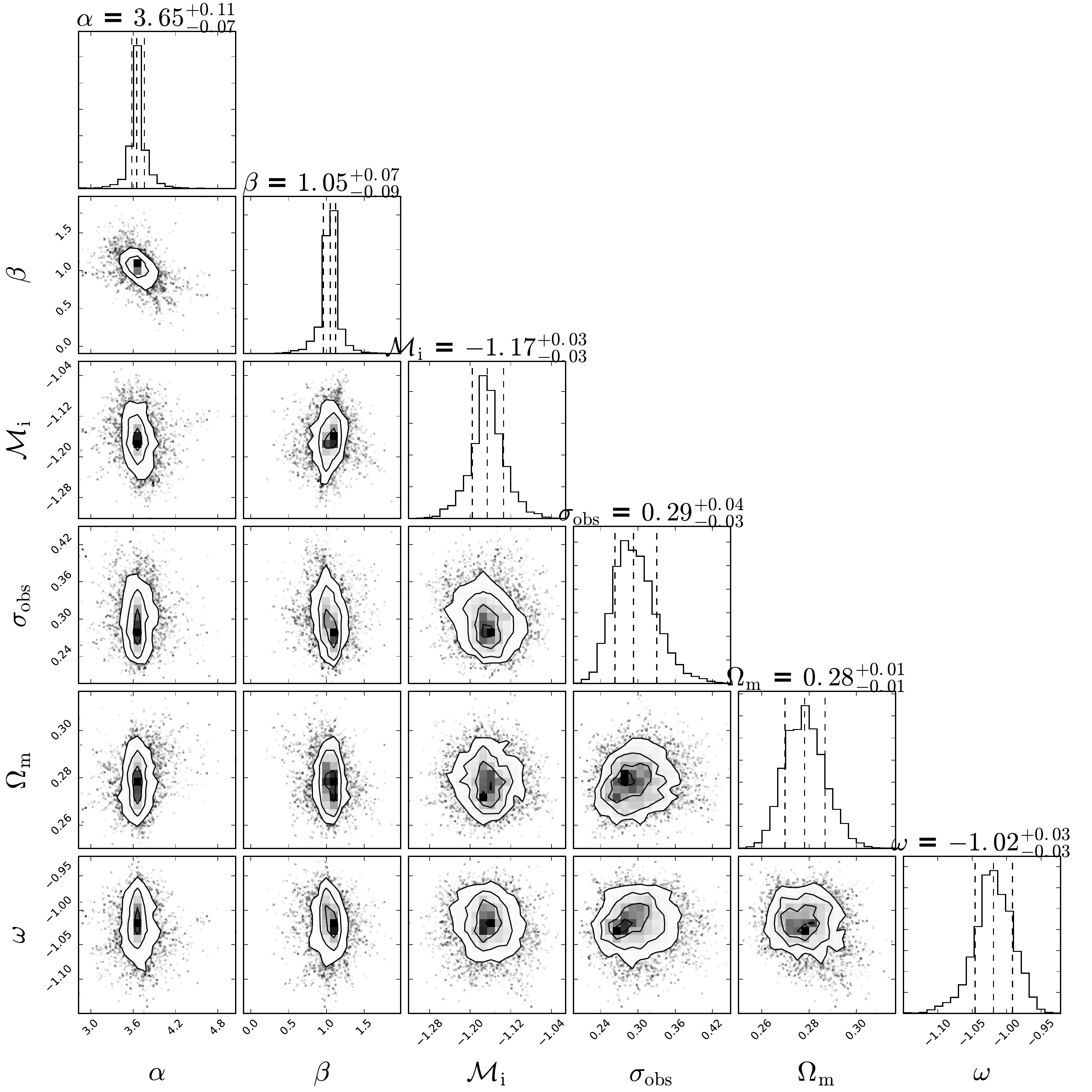}
\caption{Using a data combination of SNe II $+$ SNe Ia $+$ BAO $+$ CMB $+$ CC, we present the 1-dimensional posterior distributions on the individual parameters and $2$-dimensional marginalized contours of the $\omega$CDM model. The $68\%$ confidence ranges of different cosmological parameters are also exhibited in the corner plot.}\label{f7}
\end{figure}
\begin{figure}
\centering
\includegraphics[scale=0.4]{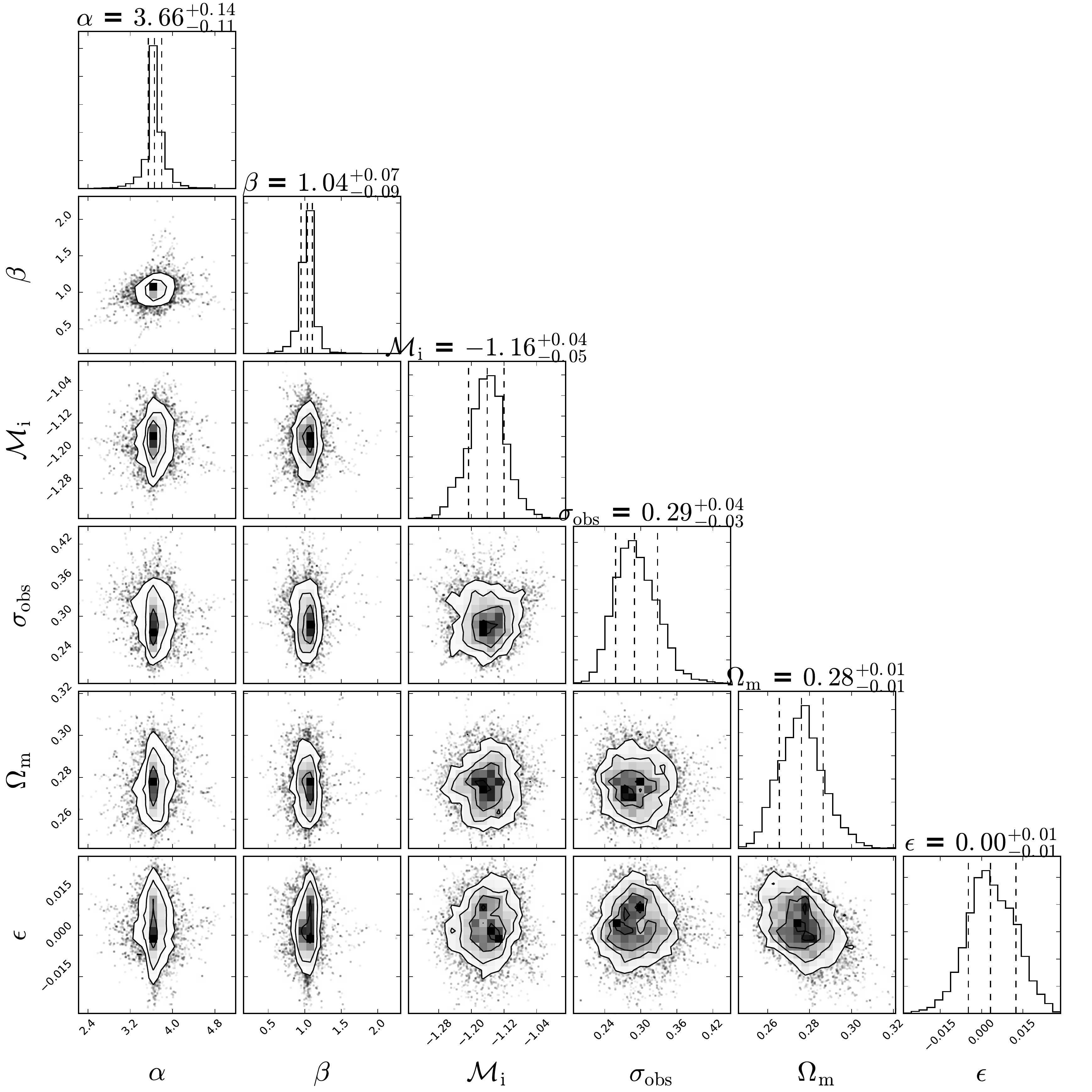}
\caption{Using a data combination of SNe II $+$ SNe Ia $+$ BAO $+$ CMB $+$ CC, we present the 1-dimensional posterior distributions on the individual parameters and $2$-dimensional marginalized contours of the DV model. The $68\%$ confidence ranges of different cosmological parameters are also exhibited in the corner plot.}\label{f8}
\end{figure}
\begin{figure}
\centering
\includegraphics[scale=0.4]{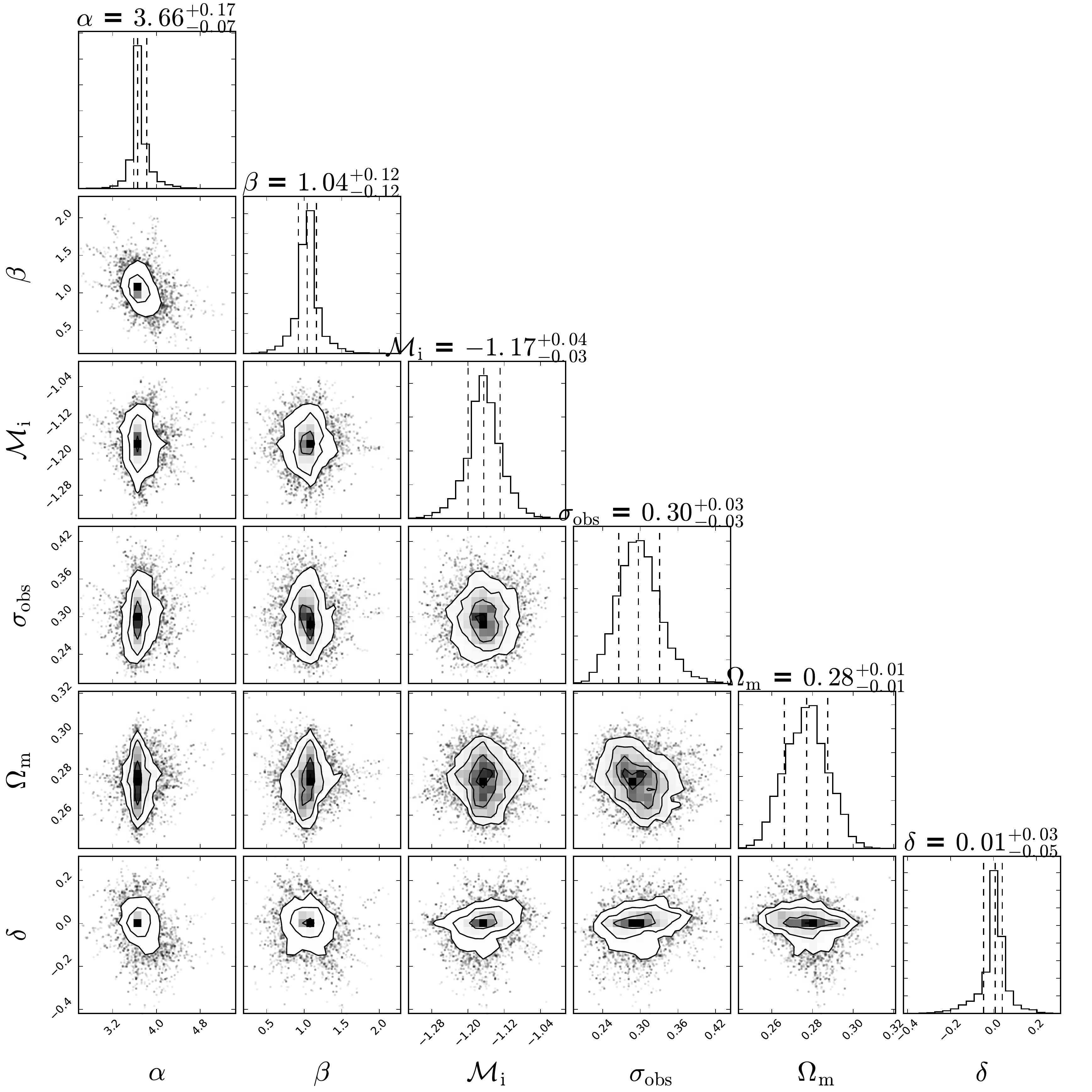}
\caption{Using a data combination of SNe II $+$ SNe Ia $+$ BAO $+$ CMB $+$ CC, we present the 1-dimensional posterior distributions on the individual parameters and $2$-dimensional marginalized contours of the DEDP model. The $68\%$ confidence ranges of different cosmological parameters are also exhibited in the corner plot.}\label{f9}
\end{figure}
Utilizing the current SNe II data to constrain four alternative cosmological models, our MCMC numerical analysis results are presented in Figs. \ref{f1}-\ref{f4},  which includes the best-fitting points and corresponding $1\sigma$ errors of individual parameters, 1-dimensional posterior distributions on the individual parameters, and 1$\sigma$, 2$\sigma$ and 3$\sigma$ $2$-dimensional marginalized contours of the o$\Lambda$CDM, $\omega$CDM, DV and DEDP models, respectively. Comparing with the constraining result of $\Lambda$CDM scenario \cite{23}, we find that the constrained values and corresponding $1\sigma$ uncertainties of $\alpha$, $\beta$, $\mathcal{M}_i$ and $\sigma_{obs}$ are very stable and independent of background cosmological models and that those of present-day matter density ratio $\Omega_m$ of the o$\Lambda$CDM, DV and DEDP models are the same with each other. Nonetheless, interestingly, the best-fitting value of $\Omega_m=0.47$ of $\omega$CDM is about 24$\%$ larger than those of other three models, and its error is still basically stable. It is noteworthy that the 1$\sigma$ uncertainties of typical parameters of above four models are all very large.

Subsequently, it is interesting to compare the constraining power from SNe II with that from the standard candles SNe Ia. Meanwhile, to implement tight constraints on the cosmological parameters, we also include BAO, CMB and CC datasets in our data analysis. We adopt the constraining result from a data combination of SNe II $+$ SNe Ia $+$ BAO $+$ CMB $+$ CC as our final result. The corresponding constraining results are presented in Figs. \ref{f5}-\ref{f9} and Tabs. \ref{t1}-\ref{t4}. One can easily find that the constraining power is not good enough by using the SNe II alone (see Tabs. \ref{t1}-\ref{t2}). If combining SNe II with SNe Ia, the constraining power can be strengthened clearly.

Furthermore, using the tightest constraint SNe II $+$ SNe Ia $+$ BAO $+$ CMB $+$ CC that we can give, we obtain the following conclusions: (i) for o$\Lambda$CDM, the constrained value $\Omega_k=-0.009\pm0.011$ is very consistent with zero cosmic curvature at the 1$\sigma$ CL, which indicates that a spatially flat universe is supported by current data in the framework of $\Lambda$CDM cosmology. Meanwhile, our result with $0.1\%$ accuracy prefers a closed universe and has the same order of magnitude with the recent Planck's restriction $|\Omega_k | < 0.005$ \cite{29}; (ii) for $\omega$CDM, the constrained EoS of DE $\omega=-1.019^{+0.028}_{-0.027}$  is very compatible with the cosmological constant scenario, which implies that there is no hint beyond the standard cosmology at the 1$\sigma$ CL; (iii) for DV, the constrained modified matter expansion rate $\epsilon=0.0032^{+0.0093}_{-0.0082}$ is also in a good agreement with zero at the 1$\sigma$ CL, which indicates that there does not exist the evidence of interaction between DM and DE in the sector of the universe; (iv) for DEDP, the constrained free parameter $\delta=0.0081^{+0.0333}_{-0.0539}$ is also well consistent with zero at the 1$\sigma$ CL, which implies that there is no hint of dynamical DE.

\section{Discussions and conclusions}
Although SNe II are fainter than SNe Ia at high redshifts, the main fact that they are more abundant than SNe Ia urge us to regard them as very useful cosmic distance indicators. With gradual SNe II data accumulation and recent several progresses \cite{21,22,23}, we believe that the SNe II cosmology will have a bright future.

Using the SCM to calibrate currently available SNe II data, for the first time, we are motivated by exploring whether there exist new physics beyond the standard cosmological model to constrain four alternative cosmological scenarios. Based on current SNe II data, we just place constraints on the above four cosmological models at the level of background evolution. From Figs. \ref{f1}-\ref{f4}, one can also find that we cannot provide tight constraints on the different typical model parameters.

To improve the constraining power further, we constrain these alternative models by using the combined datasets SNe II $+$ SNe Ia $+$ BAO $+$ CMB $+$ CC.
For all four models, we find that the constrained values and corresponding $1\sigma$ uncertainties of the intrinsic parameters $\alpha$, $\beta$, $\mathcal{M}_i$ and $\sigma_{obs}$ characterizing the evolution of SNe II are very stable and independent of background cosmological models. Meanwhile, at the 1$\sigma$ CL, we find that a spatially flat universe is preferred by current SNe II data for o$\omega$CDM, that the constrained EoS of DE is compatible with the standard cosmology for $\omega$CDM, that there is no evidence of interaction between DM and DE for DV, and that there does not exist hint of the dynamic DE.

Very interestingly, for the first time, we give the tightest constraints on the intrinsic parameters $\alpha$ and $\beta$, whose $1\sigma$ errors from a data combination of SNe II $+$ SNe Ia $+$ BAO $+$ CMB $+$ CC are at least twice smaller than those from SNe II alone (see Figs. \ref{f1}-\ref{f9}).

Notice that another important reason to use SNe II as complementary distance probes is that their progenitors and environments (only late-type galaxies) are better understood than those of SNe Ia. In future, the rapid development of SNe II cosmology will bring us more useful information about the formation and evolution of the universe.

\section*{Acknowledgements}
Deng Wang is grateful to T. de Jaeger for a long-term communication on SNe II, and thanks Yang-Jie Yan, Wei Zhang, Yuan Sun, Xin Hao, Tong Tong and Tian Tian for helpful discussions on cosmology and gravity.


\end{document}